\documentclass[conference]{IEEEtran}
\IEEEoverridecommandlockouts
\usepackage{cite}
\usepackage{amsmath,amssymb,amsfonts}
\usepackage{graphicx}
\usepackage{textcomp}
\usepackage{xcolor}
\usepackage{algorithm}
\usepackage{algorithmic}
\usepackage{mathrsfs}
\usepackage{subcaption}
\usepackage{multirow}
\usepackage{adjustbox}
\usepackage[labelformat=simple, labelsep=colon]{subcaption}

\def\BibTeX{{\rm B\kern-.05em{\sc i\kern-.025em b}\kern-.08em
    T\kern-.1667em\lower.7ex\hbox{E}\kern-.125emX}}
    
\begin{document}

\title{Blockchain-enabled Energy Trading and Battery-based Sharing in Microgrids}


\author{
    Abdulrezzak Zekiye$^\star$, Ouns Bouachir$^\dag$, \"Oznur \"Ozkasap$^\star$, Moayad Aloqaily$^\ddag$ \vspace{5px} \\
    {$^\star${Ko\c{c} University, Department of Computer Engineering, Istanbul, Turkey}}\\
    {$^\dag$College of Technological Innovation (CTI), Zayed University, UAE}\\
    {$^\ddag$Mohamed bin Zayed University of Artificial Intelligence (MBZUAI), UAE}\\
    Emails:{$^\star$\{azakieh22, oozkasap\}@ku.edu.tr, $^\dag$ouns.bouachir@zu.ac.ae, $^\ddag$moayad.aloqaily@mbzuai.ac.ae
    }
}

\IEEEoverridecommandlockouts

\maketitle
\IEEEpubidadjcol

\begin{abstract}
Carbon footprint reduction can be achieved through various methods, including the adoption of renewable energy sources. The installation of such sources, like photovoltaic panels, while environmentally beneficial, is cost-prohibitive for many. Those lacking photovoltaic solutions typically resort to purchasing energy from utility grids that often rely on fossil fuels. Moreover, when users produce their own energy, they may generate excess that goes unused, leading to inefficiencies. To address these challenges, this paper proposes innovative blockchain-enabled energy-sharing algorithms that allow consumers -without financial means- to access energy through the use of their own energy storage units. We explore two sharing models: a centralized method and a peer-to-peer (P2P) one. Our analysis reveals that the P2P model is more effective, enhancing the sharing process significantly compared to the centralized method. We also demonstrate that, when contrasted with traditional battery-supported trading algorithm, the P2P sharing algorithm substantially reduces wasted energy and energy purchases from the grid by \textcolor{black}{73.6\%}, and \textcolor{black}{12.3\%} respectively. The proposed system utilizes smart contracts to decentralize its structure, address the single point of failure concern, improve overall system transparency, and facilitate peer-to-peer payments.
\end{abstract}


\section{Introduction}
Energy is crucial for our everyday life, and the advance in technology has led to a huge increase in energy consumption \cite{energyStatistics}. Traditionally, energy is provided to consumers by a centralized utility grid, operated by a company or more. In recent years, however, consumers have been using renewable energy to decrease the cost of ordinary energy, reduce the amount of produced carbon, and depend less on the utility grid. Such consumers, who also produce energy, are often called prosumers \cite{gautier2018prosumers}. Depending on several factors such as the weather, the amount of consumed energy, and the amount of produced energy, prosumers might have an excess or shortage of generated energy at a specific time. When having an excess of energy, prosumers will be able to trade the excess amount or part of it. Trading energy can be done by selling the excess energy back to the utility grid. This approach is basically controlled by the utility grid company, where it sets the price, which is called feed-in-tariffs (FiT), and usually, it is a lower price than what they sell. It is noticeable in \cite{engAust} that in the "Solar Max \& Flexi Plan", the rate of buying energy is 37.7575 cents/kWh, while the FiT is 12 cents/kWh. In other words, the prosumer sells the energy to the grid at a price less than the price he will pay to get energy from the utility grid. That is one reason that encouraged peer-to-peer (P2P) energy trading. In peer-to-peer energy trading, prosumers sell the excess energy to other consumers at a price that both parties agree on \cite{soto2021peer}. \textcolor{black}{On the other hand, energy sharing is exchanging energy using non-monetary means. Researchers in \cite{oikonomakou2019energy, singh2018exploring, bouachir2022federatedgrids} have differentiated between the "sharing" and "trading" terms explicitly where \textbf{Energy Trading} can be defined as the exchange of energy for money and \textbf{Energy Sharing} as the exchange of energy for any type of benefit other than money.}

P2P energy trading has been handled in different ways, and one of the most recent ones is blockchain-based energy trading where the blockchain and smart contracts are used to store the data, make the system decentralized, and remove the need for third parties to operate the system, i.e. no single point of failure \cite{zekiye2023blockchain}. Blockchain is a technology based on a decentralized and distributed ledger, providing trustless, transparent, and secure storage for transactions \cite{zheng2018blockchain}. Smart contracts, executed over the blockchain, enable the enforcement of a consensus algorithm among users, as they are immutable and resistant to alteration by a single entity \cite{mohanta2018overview}. 
Blockchains exist in three types: public, consortium, and private. Public blockchains, or permissionless, are fully decentralized, allowing any user to participate without requiring permission. Consortium blockchains are controlled by a group of users, while private blockchains are managed by a single entity. Consortium and private blockchains are permissioned, necessitating authorization for participation.

In this paper, we present a permissionless blockchain-enabled micro-grid peer-to-peer energy trading system with energy sharing enabled by energy storage. The main contributions are as follows:
\begin{enumerate}
 \item Proposal of a novel method for peer-to-peer energy sharing based on consumers' energy storage. Two different sharing approaches were proposed, a centralized and peer-to-peer one.
 \item Quantitative analysis demonstrated that peer-to-peer energy sharing can reduce wasted energy and reliance on grid purchases by \textcolor{black}{73.6\%} and \textcolor{black}{12.3\%} respectively.
 \item The sharing approach demonstrated its utility, benefiting not only consumers but also prosumers who sought energy sharing when facing financial constraints.
 \item The implementation of the proposed system involved utilizing smart contracts to establish a decentralized and trustless framework.
\end{enumerate} 

This paper is organized as follows. Section \ref{rw} provides an overview of the related works. Section \ref{materials_and_methods} demonstrates the materials and methods used where we describe the generated dataset and the novel methods we provided for enabling energy sharing between peers. Section \ref{experiments} discusses the simulation details and the obtained results of the proposed novel method compared to two different scenarios. Finally, \textcolor{black}{Section \ref{discussion} discusses the proposed systems, and} Section \ref{conc} states the conclusions.

\section{Related Works} \label{rw}





Peer-to-peer trading energy has \textcolor{black}{received significant attention} in the literature \cite{  long2017peer, perk2020joulin, aloqaily2022synergygrids}. However, research regarding sharing energy is limited. \textcolor{black}{Researchers in \cite{singh2018exploring} conducted an ethnographic study conducted at two off-grid villages over 11 months, focusing on returns received by residents. They identified three types of returns: in-cash, in-kind, and intangible. In their work, in-cash denotes monetary payment, in-kind signifies payment through goods or services, and intangible return encompasses unmeasured and unquantified social gestures and actions.} FederatedGrids introduced sharing energy between peers where the sharing phase is handled after the trading phase \cite{bouachir2022federatedgrids}. Sharing requests are placed by consumers with no funds and those requests are handled by prosumers with excess energy that was not sold. They mentioned that the sharing is for future benefits to be provided to the prosumer; however, they have not implemented such future benefits. In fact, such benefits need to be discussed in detail to prevent greedy consumers from getting energy for free without providing any benefit in return. 

\textcolor{black}{The utilization of energy storage in energy trading systems has been explored in the literature, as evidenced by \cite{tushar2016energy, liu2017cloud, he2021peer}. In \cite{tushar2016energy}, batteries at peers were controlled by another entity, proposing a joint energy storage ownership scheme where residential units lease part of their energy storage to shared facility controllers. However, this approach does not enable peers to directly benefit from each other's batteries.}



Researchers in \cite{liu2017cloud} suggested the use of centralized energy storage (CES) devices to offer storage services to distributed users. This enables residential energy consumers and commercial users to rent CES power and capacity, resulting in reduced personal energy expenditures. 

Finally, in \cite{he2021peer}, a peer-to-peer (P2P) energy trading framework for distributed photovoltaic (PV) prosumers and consumers was proposed. The framework establishes a community-sharing market facilitated by an Energy Provider (EP), who installs, manages, and maintains a publicly accessible battery energy storage (ES) system. \textcolor{black}{In other words, the peers in their proposed system benefit from a central energy storage.}

Our proposed solution addresses the research gap by introducing a method that enables peers to utilize each other's energy storage in exchange for shared energy. This approach maximizes the utilization of energy storage and provides a means for consumers with insufficient funds to access energy using their energy storage. \textcolor{black}{Compared to \cite{bouachir2022federatedgrids}, the problem of greedy consumers was solved by asking the consumers (who would like to get energy without being able to pay for it) to \textcolor{black}{share} part of their energy storage for a specific amount of time, in exchange for the provided energy. In addition, the peers in our proposed system can benefit from other peers' energy storage, unlike \cite{tushar2016energy,  liu2017cloud, he2021peer} where either a central entity benefits from peers' batteries, or the peers benefit from a central battery.}


\begin{table}[h]
\caption{Notation used in the equations and algorithms}
\centering
\begin{adjustbox}{width=1\columnwidth,center}
\begin{tabular}{|c|l|}
\hline
\textbf{Symbol} & \textbf{Explanation}           \\ \hline
$p_t$         & P2P trading price at time step t  (EUR) \\ \hline
$up_t$         & Utility grid price at time step t (EUR) \\ \hline
$R_t$             & Number of energy requests in time step t                 \\ \hline
$O_t$         & Number of energy offers in time step t  \\ \hline
$hEDGW$         & Hourly energy demand generation and weather dataset  \\ \hline
$hEDGW(x)$         & Column x of Hourly energy demand generation and weather dataset  \\ \hline
$hEDGW(x, t)$  & Column x of Hourly energy demand generation and weather dataset \\ & at time step t \\ \hline
$\eta$         & The usable percentage of shared energy  \\ \hline
$\tau$         & The expiry time for the shared energy stored in consumer's energy storage\\ \hline
$h$         & Set of N houses in microgrid\\ \hline
$h_i$         & House \textit{i} in microgrid\\ \hline

$h_i.ee$         & Excess energy of house \textit{i} in microgrid. A negative value indicates a need for energy.\\ \hline
$h_i.rc$         & Remaining capacity of house \textit{i}'s energy storage\\ \hline
$h_i.re$         & Reserved energy stored in house \textit{i}'s energy storage\\ \hline
$h_i.balance$         & Balance of house \textit{i}\\ \hline
$h_i.b$ & Battery of house \textit{i}\\ \hline

$sm$ & \textcolor{black}{The smart contract}\\ \hline
$sm.balance$ & \textcolor{black}{The balance of the smart contract}\\ \hline
\end{tabular}
\label{t1}
\end{adjustbox}
\end{table}

\section{Material and Methods} \label{materials_and_methods}

\subsection{Dataset}\label{dataset}
For the proposed peer-to-peer energy trading and sharing system that depends on energy storage, relevant data per entity is needed in order to simulate the system and evaluate its efficiency. The relevant needed data are consumption and production amounts, \textcolor{black}{energy} price, and energy storage data (battery capacity and charge). According to our knowledge, there is no such dataset. For that reason, we generated the required data to be used in the simulation. Several steps were followed to make the generated data as close as possible to real-life data. First, we used the Hourly Energy Demand Generation and Weather dataset \cite{jhana2019hourly} to generate the load or consumption data per house, in addition to getting the real price per watt. Regarding the production data per house, each house has \textcolor{black}{a probability ($pr$)} of having solar panels or not\textcolor{black}{.} The amount of generated energy at each house depends on the weather, the time of the day, and the number of solar panels at each house, where the output of a solar panel is between 170 and 350 watts per hour \cite{solarPanelPower}. According to \cite{numberOfPanels}, the average number of needed solar panels per house is between 17 and 21. For each house, the number of solar panels and the output of each one were generated randomly according to the previously mentioned ranges. As a result, the produced energy for a house with index \(i\) is calculated as in \eqref{eq:gsp} \textcolor{black}{with a probability $pr$}. However, this equation does not reflect the weather data or the time of day. Hourly Energy Demand Generation and Weather dataset has a column representing the hourly real generated energy from solar panels in Spain, which we used to make our synthesized "produced energy" column reflects real-life data, as a result, it will naturally take into consideration the weather and the time of the day. To achieve that, for each time step, we multiplied the calculated produced energy from Equ.\eqref{eq:gsp} by the corresponding produced solar energy from the Hourly energy demand generation and weather, divided by its maximum value. Thus, the produced solar energy for house \(i\) at time step \(t\) is calculated using Equ.\eqref{eq:gsph}. The energy storage at each house was generated randomly between 5 and 15 kilowatt-hour (kWh).

\begin{equation}
\begin{aligned}
\text{generated solar power}(i) = {} & \text{random}(17, 20) *\\ 
                    & \text{random}(170, 350)
\end{aligned}
\label{eq:gsp}
\end{equation}

\begin{equation}
\begin{aligned}
\text{generated solar power}(i,t) = {} \\ \text{generated solar power}(i) * \\ \frac{\text{hEDGW}('generation\ solar',t)}{\text{max}(\text{hEDGW}('generation\ solar'))}  
\end{aligned}
\label{eq:gsph}
\end{equation}

\subsection{Pricing}\label{sec:pricing}
Pricing is an extensively explored topic in the literature \cite{liu2018dynamic,alsalloum2020systematical}. Within our system, we adopt a simple microgrid-fixed pricing technique, wherein the price is determined based on recent prices and incorporates the count of energy offers and requests.
In scenarios where the number of requests surpasses the number of offers, prices would rise, and conversely, if the number of offers exceeds the number of requests, prices would decrease. The count of requests signifies the number of consumers willing to pay, while the count of offers represents the current prosumer-made offers. It is essential to ensure that the price of energy within the microgrid remains below the utility grid price \textcolor{black}{and above the FiT price} to maintain the viability of the trading and sharing concept. Consequently, the pricing equation is formulated as depicted in Equ.\eqref{eq:pricing}.

\begin{equation}
\begin{aligned}
\text{p}_t = \max(FiT, \min(\text{up}_t, &  \frac{R_t}{\text{O}_t} \cdot \frac{\text{p}_{t - 1} + \text{p}_{t - 2} + \text{p}_{t - 3}}{3}))
\end{aligned}
\label{eq:pricing}
\end{equation}

\(\text{p}_t\) is the calculated price at the current time step \(t\). \(\text{up}_t\) is the utility price at time step \(t\). \(R_t\), and \(O_t\) are the number of trading requests by consumers and the number of offers made by prosumers at time step \(t\) respectively.

\subsection{Energy Trading and Sharing}
\begin{figure}
    \centering
    \includegraphics[width=0.54\linewidth]{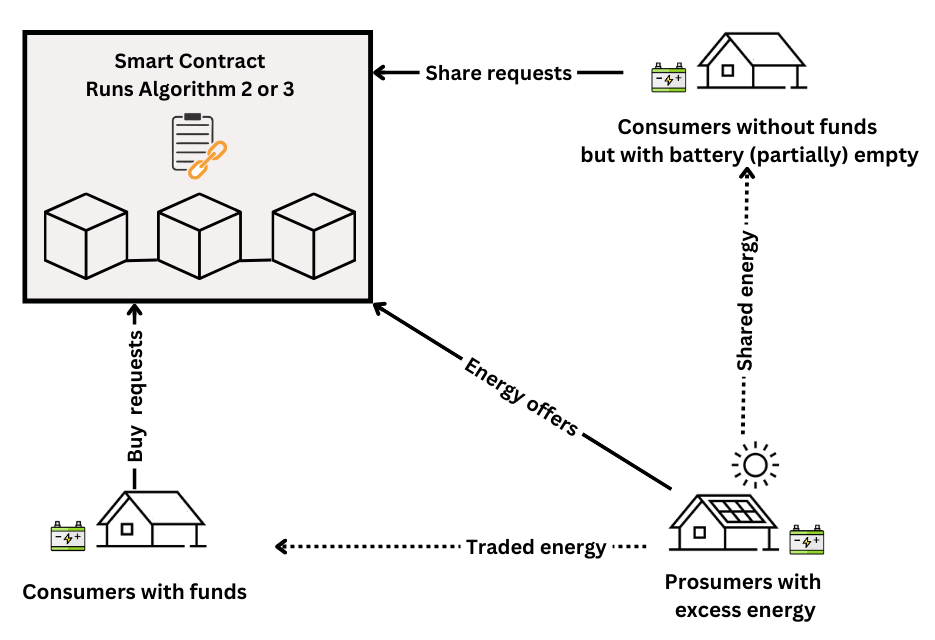}
    \caption{Proposed system overview}
    \label{fig:sys}
\end{figure}

The proposed system, Blockchain-enabled Energy Trading and Battery-based Sharing in Microgrids, involves prosumers and consumers as distinct entities. Both prosumers and consumers are equipped with energy storage units (batteries) that vary in capacities. Prosumers, possessing surplus energy, are willing to sell, while consumers require energy either due to no generated energy at all or an energy demand surpassing the generated and stored energy. \textcolor{black}{An overview of the system is depicted in Figure \ref{fig:sys}}.

The proposed system emphasizes sharing energy in exchange for utilizing the consumer's energy storage, essentially employing the energy storage as a form of payment. Initially, the system engages in energy trading between prosumers and consumers with available funds, utilizing a first-come, first-served matching process due to uniform prices for all offers and requests. Subsequent to the trading phase, if there are remaining unmatched energy offers and there are consumers who lack funds but have partially empty energy storage, the system facilitates energy sharing.

With the support of a sharer entity (${SE}$), energy sharing occurs and the shared energy is transferred from the prosumer to the consumer. The consumer utilizes a percentage of the shared energy ($\eta$) for his own needs, while the remaining energy is stored in his energy storage for a specified time ($\tau$). The sharer entity (${SE}$) has the capability to sell this stored energy from the consumer's energy storage within the designated expiry time ($\tau$). After $\tau$ time steps, the sharer entity loses the ability to sell the stored energy, and the consumer can use it for his own purposes. The reason beyond introducing an expiry time ($\tau$) is to not occupy the consumer's battery for a long time, which might be indefinite. We explore and simulate two distinct sharer entities that enable sharing: (i) a centralized approach and (ii) a peer-to-peer approach. Prior to trading or sharing, all peers submit either an energy offer, a buying request, or a sharing request according to Algorithm \ref{alg:submit}.

\subsubsection{Centralized Sharer Entity: C-SE}

Algorithm \ref{alg:centralized_sharing} illustrates the centralized sharer approach, where a third entity is assumed to collect fees from each energy trade transaction. These fees are utilized to facilitate energy sharing, with the entity paying the prosumer wishing to share energy and transferring the shared energy to the consumer. The stored energy in the consumer's energy storage is then sold by this third entity within ($\tau$) time steps. In this approach, the third entity can be conceptualized as an investor seeking income by supporting the sharing processes. The smart contract that runs the proposed system is the third party in our system. In other words, the fees will be transferred to the smart contract balance, the smart contract will be paying prosumers for their shared energy, and it will receive payments upon selling the stored energy in the consumer's energy storage.

\subsubsection{Peer-to-peer Sharer Entity: P2P-SE}
In this approach, the prosumer with an unmatched excess of energy becomes the sharer entity for his own energy, as illustrated in Algorithm \ref{alg:p2psharing}. That is to say, the prosumer will share energy with the consumer first \textcolor{black}{where the consumer will be able to use $\eta$\% of it directly}. Then, the prosumer will be able to sell (1- $\eta$)\% of the shared energy from the consumer's energy storage within the next $\tau$ time steps. \textcolor{black}{As in the C-SE, after $\tau$ time steps, the consumer can utilize what remains from the (1- $\eta$)\% of the shared energy in their energy storage, if any is left.}

In both systems, \textcolor{black}{Algorithm \ref{alg:submit} will be executed first. In this algorithm,} consumers and prosumers have energy storage that will be charged/discharged before making any energy offer/request. In other words, prosumers will charge their batteries first, and if there is still an excess of energy, an energy offer will be made. Similarly, consumers will discharge their batteries first, and if there is still a need for energy, either a trading request or a sharing request will take place. A trading request will occur when the consumer has funds to pay for the requested energy. If the consumer's funds are insufficient, a share request will be made in case there is a part of the energy storage empty. During each time step, the price will be calculated and fixed for that time step, as described in Section \ref{sec:pricing}. \textcolor{black}{In Algorithms \ref{alg:centralized_sharing} and \ref{alg:p2psharing}, the match function represents a first-come, first-served matching process.}


\begin{algorithm}[t]
\caption{Submitting energy offer, buying request, or sharing request}
\begin{algorithmic}[1]
\label{alg:submit}
\STATE \textbf{input:} \textit{h $\leftarrow$ $[h_1,..., h_n]$, $expectedPrice$}
\STATE \textbf{output:} \textit{offers, buyRequests, sharingRequests}

\FOR{$ h_i  \in  h $}
\IF{$ h_i.ee > 0 $ }

\STATE $chargedAmount = min(h_i.ee, h_i.rc)$
\STATE chargeBattery($h_i.b$, $chargedAmount$)

\STATE $h_i.ee = h_i.ee - chargedAmount$
\IF{$ h_i.ee > 0 $ }
\STATE $offers.append($prosumer=$h_i$, $energy\_amount$=$h_i.ee$)
\ENDIF
\ELSIF{$ h_i.ee < 0 $}
\STATE $chargeUsed = min(h_i.ee, h_i.bc - h_i.re)$
\STATE $h_i.ee = h_i.ee + chargeUsed$
    \IF{$ h_i.ee < 0 $ }
        \IF{$ h_i.balance > h_i.ee * expectedPrice $ }
            \STATE $buyRequests.append($consumer=$h_i$, energy\_amount=$h_i.ee$)
        \ELSIF{$h_i.rc > 0$}
            \STATE $requestedAmount = min(-1 * h_i.ee , h_i.rc)$
            \STATE $shareRequests.append($consumer=$h_i$, requested\_amount=$requestedAmount$ )
        \ENDIF
    \ENDIF
\ENDIF

\ENDFOR
\end{algorithmic}
\label{alg:submit}
\end{algorithm}

\begin{algorithm}[t]
\caption{Matching energy offers with buying and sharing requests in C-SE}
\begin{algorithmic}[1]
\STATE \textbf{input:} \textit{offers, buyRequests, sharingRequests, sharedEnergies, $\eta$, $\tau$, fee, $p_t$}
\STATE $shared\_energies\_\tau$ =  sharedEnergies within previously $\tau$ timesteps
\STATE match($buyRequests$, $shared\_energies\_\tau$) 
\STATE match($buyRequests$, $offers$) 

\IF{offers are not all matched}
    \IF{sharingRequests $\neq \emptyset$}
        \FOR{$ offer  \in  offers $}
            \FOR{$ sh  \in  sharingRequests $}
                \STATE $energy = min(offer.amount, sh.amount)$
                
                \IF{$ sm.balance > energy * p_t $}
                    \STATE ($offer.prosumer$) gets paid ($energy * p_t$) by smart contract after deducting fee\%
                    \STATE                 $sharedEnergies.append$($seller = smart\ contract,amount= energy$ * $(1 - \eta)$)

                    \STATE store $energy * (1 - \eta)$ in consumer's battery
                    \STATE transfer $energy * (\eta)$ to consumer to use
                \ENDIF
            \ENDFOR
        \ENDFOR
    \ENDIF
\ENDIF
\end{algorithmic}
\label{alg:centralized_sharing}
\end{algorithm}

\begin{algorithm}[t]
\caption{Matching energy offers with buying and sharing requests in P2P-SE}
\begin{algorithmic}[1]
\STATE \textbf{input:} \textit{offers, buyRequests, sharingRequests, sharedEnergies, $\eta$, $\tau$, $p_t$}

\STATE $shared\_energies\_\tau$ =  sharedEnergies within previously $\tau$ timesteps
\STATE match($buyRequests$, $shared\_energies\_\tau$) 
\STATE match($buyRequests$, $offers$) 
\IF{offers are not all matched}
    \IF{sharingRequests $\neq \emptyset$}
        \FOR{$ offer  \in  offers $}
            \FOR{$ sh  \in  sharingRequests $}
                \STATE $energy = min(offer.amount, sh.amount)$
\STATE                 $sharedEnergies.append$($seller = offer.prosumer,amount= energy$ * $(1 - \eta)$)
                \STATE store $energy * (1 - \eta)$ in consumer's battery
                \STATE transfer $energy * \eta$ to consumer to use
            \ENDFOR
        \ENDFOR
    \ENDIF
\ENDIF
\end{algorithmic}
\label{alg:p2psharing}
\end{algorithm}

\section{Experiments and Results} \label{experiments}
The proposed system has been implemented using Solidity and can be deployed to any Ethereum-compatible blockchain, \textcolor{black}{where we deployed it on Sepolia test net. The role of the smart contract is to match the offers and previously shared energy with energy requests in a decentralized and transparent manner. In other words, the smart contract will be running Algorithm \ref{alg:centralized_sharing} or \ref{alg:p2psharing}, where it also plays the role of the sharer entity in the C-SE case}. 
Simulation of the system has been done using Python. For the purpose of simplification, the gas fees of interacting with the smart contract were not simulated. We compared the results of the proposed sharing scenarios to each other and to the following scenarios:

\noindent\textbf{Trading:} Trading energy without energy storage installed at peers.
    
\noindent\textbf{Trading with Batteries (T\&B):} Trading energy between peers with energy storage installed at each peer. The energy storage is charged/discharged before selling/buying energy, and each peer utilizes only his energy storage.
    
\begin{figure}[t]
  \centering
  \begin{subfigure}[b]{0.96\columnwidth}
    \includegraphics[width=\columnwidth]{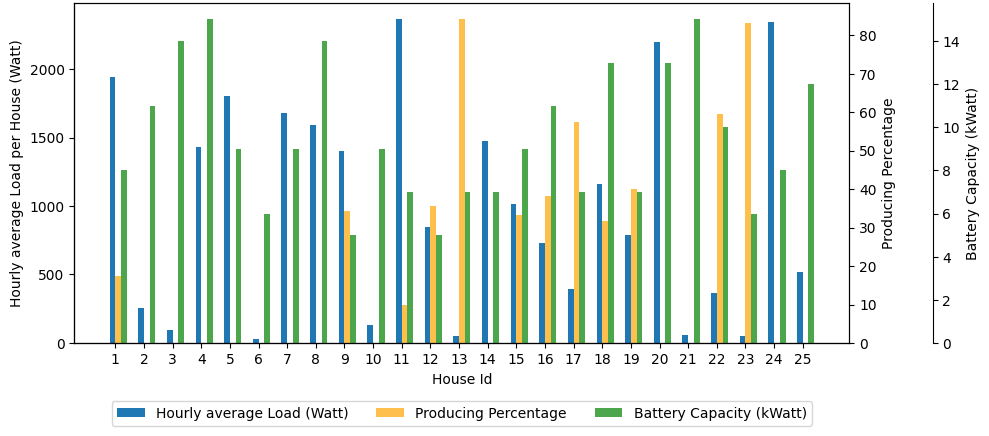}
    \caption{}
    \label{fig:dataset}
  \end{subfigure}
  \begin{subfigure}[b]{0.81\columnwidth}
    \includegraphics[width=\columnwidth]{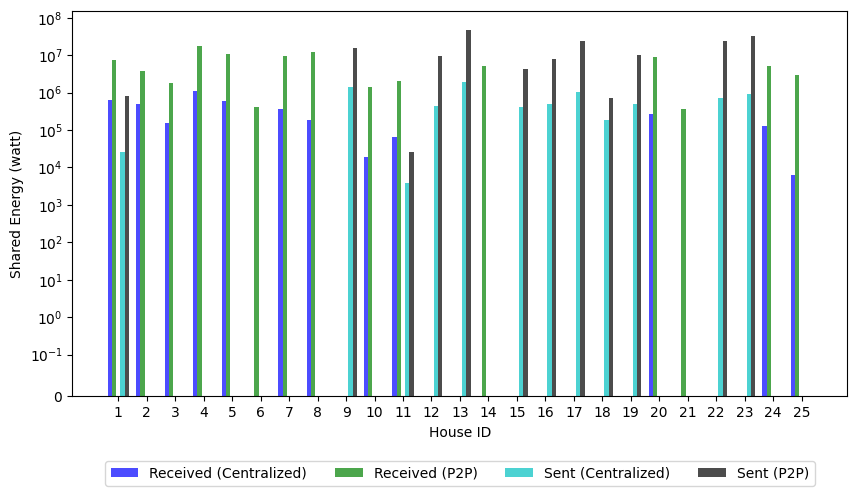}
    \caption{}
    \label{fig:p2p-sharing-per-house}
  \end{subfigure}
  \caption{(a) Producing percentage, \textcolor{black}{average hourly load}, and battery capacity for each house in the \textcolor{black}{balanced} dataset, (b) Comparison of shared energy amounts: C-SE and P2P-SE}
  \label{fig:whole}
\end{figure}

The number of simulated houses is 25 with details as shown in Figure \ref{fig:dataset} where the producing percentage represents the timeslots percentage the house was producing energy (0 means the house is a consumer all the time). \textcolor{black}{The dataset was generated with a 0.5 probability of being a prosumer, resulting in 13 consumers and 12 prosumers. The total energy generated in this experiment accounted for approximately 58\% of the total required energy}. Each house started the simulation with a balance of 100 Euros. Table \ref{tab:results} shows \textcolor{black}{the results of the simulation for the aforementioned trading/sharing algorithms}. The simulation assumes that energy generation, consumption, and utility price are known at the time step at which the trading or sharing will happen. The expiry time for the shared energy stored in consumer’s energy storage was set to 12 time steps (12 hours) ($\tau$ = 12) and the usable percentage of shared energy was set to 50\% ($\eta$ = 0.5) in both sharing scenarios. \textcolor{black}{Finally, FiT was estimated as one-third of the utility grid price.}

\subsection{Centralized Sharer Entity: C-SE}
In the C-SE case, the smart contract acts as the sharer entity. We assumed that the fee on any selling transaction is 10\%. 
At the end of the simulation, the smart contract made sales totaling \textcolor{black}{168.89} Euros and accrued \textcolor{black}{181} Euros in fees. \textcolor{black}{Despite these earnings, it disbursed} \textcolor{black}{349.89} Euros to the sharers, so it lost at the end of the simulation, or it did not earn anything. We can see that the fees are what enable this sharing scenario, in other words, the sharing is being supported by the users themselves. More specifically, prosumers will be receiving less money as a result of the fees. That loss came because of each shared energy the contract pays for, half of it ($\eta$) will be saleable by the contract, which means it has to sell it at least double the price to at least not suffer a loss. In addition, the saleable energy can be sold within only the expiry date, which means it will lose the energy after that expiry date if it does not sell it. Compared to trading with batteries (T\&B) approach, the total wasted energy is reduced by \textcolor{black}{2.23\%} approximately. Also, the dependence on the utility grid has been reduced by \textcolor{black}{0.17\%}, approximately.

\subsection{Peer-to-peer Sharer Entity: P2P-SE}
In this scenario, there is no third entity that enables the sharing process and the prosumers themselves will be responsible for selling the percentage of shared energy (1-$\eta$) within $\tau$ timesteps. By comparing P2P-SE to T\&B, it is noticeable from Table \ref{tab:results} that P2P-SE decreased wasted energy by \textcolor{black}{73.6\%} and the dependency on the utility grid by \textcolor{black}{12.3\%}. 




\begin{table*}[h]
\centering

\caption{Comparison of trading/sharing scenarios}
\begin{tabular}{|l|c|c|c|c|c|c|}
\hline
\textbf{Metric} &\textbf{ No Trading} & \textbf{Trading} & \textbf{T\&B} & \textbf{C-SE }& \textbf{P2P-SE}\\
\hline
Total Energy from Grid (watt) & 677,245,700 & 636,830,500 & 590,534,700	 & 589,513,400 & 512,560,500  \\ \hline
Total Paid to Grid (EUR)& 39,710.62 & 37,295 & 34,580.56 & 34,512.42 & 30,321.51\\ \hline

\textcolor{black}{Total Earned from P2P Trading (EUR)} & 0.0 & 2,015.87	 & 1,595.36 & 1,460.19	 & 1,110.92 \\ \hline
Total Energy Wasted (watt) & 313824966 & 277632210 & 225543162 & 217701915 & 55616664 \\ \hline
Total Shared by Prosumers (watt) & 0.0 & 0.0 & 0.0 & 8,100,853 & 178,308,500  \\ \hline
Total Earned from Sharing (EUR) & 0.0 & 0.0 & 0.0 & 349.89 & 667.63  \\ 
\hline
\end{tabular}
\label{tab:results}
\end{table*}

Figure \ref{fig:p2p-sharing-per-house} presents a comparative analysis of shared energy amounts in C-SE and P2P-SE scenarios for each house. Notably, P2P-SE exhibits a higher quantity of shared energy compared to C-SE, marking a substantial increase in the total shared energy by \textcolor{black}{2101.11}\%. Furthermore, it is evident that both consumers and prosumers benefit from the shared energy. Specifically, prosumers numbered 1 and 11 actively engaged in both sending and receiving shared energy, while other \textcolor{black}{prosumers} exclusively sent shared energy. It is noticeable that consumers received shared energy which means they reached a timestep where they had no funds, yet they were able to get energy. \textcolor{black}{Nevertheless, it is noteworthy that consumers 6, 14, and 21 exclusively received shared energy within the P2P-SE scenario only, underscoring the superiority of P2P-SE over C-SE.}  In conclusion, energy sharing proves advantageous for both prosumers and consumers, with prosumers occasionally requesting shared energy based on their energy needs.

 
\section{Discussion} \label{discussion}
In the results presented earlier, it was observed that prosumers generated approximately 58\% of the required energy. Subsequently, the proposed system underwent testing using two additional scenarios. In the first case, prosumers produced approximately 12\% of the required energy, while in the second case, they produced around 128\% of the required energy (total generated energy is 28\% more than the total needed energy). \textcolor{black}{When energy production amounted to 12\% of the total, no sharing occurred among prosumers, who utilized and traded all of it. However, when energy production was 128\% of the required amount, sharing occurred in both C-SE and P2P-SE scenarios, reducing grid dependency by 0.31\% and 5.2\% for C-SE and P2P-SE, respectively.}


Comparing C-SE and P2P-SE, it is shown that P2P-SE outperforms C-SE in various aspects: enhancing shared energy, boosting prosumer profits through reselling the energy they have stored on consumers' energy storage, and reducing reliance on the utility grid. Nevertheless, within the P2P-SE algorithm, prosumers share energy with the expectation of selling it within the subsequent $\tau$ time steps. Yet, there exist scenarios where certain prosumers may fail to offload this energy. In such instances, C-SE might offer a more favorable option for these prosumers. Identifying these specific cases and devising an algorithm capable of dynamically sharing energy using either C-SE or P2P-SE represents a future objective.


\section{Conclusions} \label{conc}
Our proposed blockchain-enabled microgrid system for energy trading and battery-based sharing involves prosumers and consumers with energy storage units, enabling energy sharing by the usage of energy storage. Simulations of centralized and peer-to-peer sharer entities were conducted on a dataset representing 25 houses. While C-SE faced financial losses, it showcased a reduction in wasted energy. Conversely, P2P-SE demonstrated significant decreases in wasted energy and dependency on the utility grid. Comparing P2P-SE to C-SE, we noticed that P2P-SE decreased the wasted energy and dependency on the utility grid by \textcolor{black}{73.02}\% and \textcolor{black}{13.05}\% respectively. These findings underscore the potential efficiency improvements and resource optimization offered by our proposed system. In the future, we plan to investigate the effects of (a) the parameters $\tau$ and $\eta$, (b) the quantity of produced energy, and (c) energy storage capacities on the system. Additionally, we aim at implementing improved pricing techniques and scale the system to operate between microgrids.

\section*{Acknowledgment}
This work is supported by TUBITAK (The Scientific and Technical Research Council of Türkiye) 2247-A National Leader Researchers Award 121C338.

\bibliographystyle{IEEEtran}
\bibliography{IEEEabrv,references}

\end{document}